\documentclass[aps, 10pt, prc, twocolumn, superscriptaddress, nofootinbib, showpacs, floatfix, longbibliography]{revtex4-2}
\usepackage{graphicx}  %for graphics
\usepackage{dcolumn}   %for decimal-point aligned tables
\usepackage{amsmath}  
\usepackage{amssymb}
\usepackage{color}  
\usepackage{orcidlink}
\usepackage{braket}
\usepackage[utf8]{inputenc}

\usepackage{hyperref}
\hypersetup{breaklinks=true,colorlinks=true,linkcolor=blue,citecolor=blue,filecolor=magenta,urlcolor=cyan}

\usepackage[all]{hypcap}

\renewcommand{\vec}[1]{\mbox{\boldmath $#1$}}

\newcommand{\sfrac}[2]{{\textstyle{\frac{#1}{#2}}}}
\begin{document}
\title{Finite-range pairing in nuclear density functional theory}
\author{Sudhanva Lalit~\orcidlink{https://orcid.org/0000-0001-7758-492X}}
\affiliation{%
 Facility for Rare Isotope Beams, Michigan State
  University, East Lansing, MI 48824, USA
}%
\author{Paul-Gerhard Reinhard~\orcidlink{0000-0002-4505-1552}}
\affiliation{Institut für Theoretische Physik II, Universität Erlangen-Nürnberg, 91058 Erlangen, Germany}
\author{Kyle Godbey~\orcidlink{0000-0003-0622-3646}}
\affiliation{%
 Facility for Rare Isotope Beams, Michigan State
  University, East Lansing, MI 48824, USA
}%
\author{Witold Nazarewicz~\orcidlink{0000-0002-8084-7425}}
\affiliation{%
 Facility for Rare Isotope Beams, Michigan State
  University, East Lansing, MI 48824, USA
}%
\affiliation{Department of Physics and Astronomy, Michigan State
  University, East Lansing, Michigan 48824, USA}

\begin{abstract}
Pairing correlations are ubiquitous in low-energy states of atomic nuclei. To incorporate them within nuclear density functional theory, 
 often used for global computations of nuclear properties,
pairing functionals that generate nucleonic pair densities and pairing fields are introduced. Many pairing functionals currently used can be traced back to zero-range nucleon-nucleon interactions. Unfortunately, such
functionals are plagued by deficiencies that become apparent in large model spaces that contain unbound single-particle (continuum) states. In particular, the underlying computational schemes diverge as the single-particle space increases, and the results depend on how marginally occupied states are incorporated. These problems become more pronounced for pairing functionals that contain gradient-density dependence, such as in the Fayans functional.
To remedy this, 
finite-range pairing functionals are introduced.
In this study, this is done by folding the pair density with Gaussians.
We show that a folding radius of about 1\,fm offers the best compromise between quality and stability, and substantially reduces the pathological behavior in different numerical applications.
\end{abstract}
\maketitle

\section{Introduction}

Shortly after the development of the Bardeen\-–Cooper\-–Schrieffer (BCS) theory of electronic  superconductivity~\cite{BCS1957}, nucleonic pairing was introduced
\cite{Bohr1958,Ring1980,Dean2003,Broglia2013}. 
Early pairing models employed a very
simple ansatz for the pairing functional, such as the constant-gap or the
constant-force monopole pairing model. With the advent of nuclear density functional theory (DFT) \cite{Bender2003,Schunck2019DFT}, local density-dependent
pairing functionals were proposed based on a zero-range pairing interaction
\cite{Saperstein1965,Bochnacki1967,Chasman1976,Kadmenski1978,Krieger1990,Fayans1994a,Fayans1996,Garrido1999,Dobaczewski2001}. In the following, we shall refer to such pairing functionals as  ``zero-range pairing functionals.''

While zero-range pairing functionals offer significant computational advantages, they give rise to ultraviolet divergences \cite{Bruun1999,Bulgac2002}. 
Thus, zero-range pairing functionals come
with a recipe for a high-momentum pairing cutoff, resulting in an effective pairing renormalization \cite{Dobaczewski1996,Niksic2005}.  An alternative to cope with the divergence
is the pairing regularization developed in analogy to well-established
regularization schemes in field theory \cite{Bulgac2002,Yu2003,Borycki2006,Pei2011,Pei2015,Dobaczewski2013}.
Although formally appealing, pairing regularization is plagued by problems similar to standard pairing cutoff schemes.  Even worse, pairing functionals
with gradient terms, as in the Fayans functionals
\cite{Fayans1994a,Fayans1996,Reinhard2017}, exhibit higher-order divergences which
would require higher-order regularization.  Particular problems
arise as soon as high-energy continuum states are present in the Hartree-Fock-Bogoliuybov (HFB) scheme. 

Most HFB solvers rely on a finite basis representation, thus
delivering a discretized approximation to the continuum. The structure
of such a pseudo-continuum depends on the actual numerical representation
(harmonic oscillator basis or coordinate-space grid, symmetry restrictions),
numerical parameters \cite{Borycki2006,Pei2008,Pei2011,Chen2022},
and beyond mean-field corrections. As a
consequence, pairing functionals are not fully portable between different HFB implementations.  Particularly
sensitive to various approximations are the BCS calculations. The HFB pairing is a bit more robust
because of the localization of canonical states \cite{Dobaczewski1996}.
This calls for pairing functionals that
are intrinsically convergent. One possible solution is to replace
zero-range functionals with finite-range ones.
Finite-range pairing functionals have already been used 
in nuclear DFT calculations. A commonly used D1S Gogny functional \cite{Berger1984,Grasso2001,Grasso2002} is a good example. A simplification of
the Gogny pairing functional, which is particularly useful for numerical applications, is the separable pairing  functional \cite{Duguet2004,Tian2009a,Tian2009,Ma2010,Vesely2011,Teeti2021}
which also involves a finite-range pairing. The present work continues along these lines.

This study explores separable pairing functionals in connection
with non-relativistic nuclear DFT models, the widely used Skyrme
functional, and, more importantly,  the Fayans functional. 
The paper is structured as follows. 
In Sec. \ref{sec:revcutoff}, we briefly review the limitations of the zero-range pairing functionals.
Section~\ref{sec:finrange} introduces the finite-range pairing
functional in the form used here. Results are presented in 
Sec.~\ref{sec:results}, which illustrates the performance of the
new functional and 
presents the results of large-scale calibration 
to nuclear ground-state data. Finally, Sec .~\ref {Conc} contains the conclusions of this work.

\section{Zero-range pairing and continuum effects}
\label{sec:revcutoff}

In this section, we discuss the effects of the continuum space in the standard BCS and HFB treatment of pairing with zero-range functionals. To limit the space of single-particle (s.p.) states used in the pairing equations, we employ here and throughout this paper a soft cutoff regulated by the s.p. energies, see
Appendix~\ref{sec:softcut} for details.

\subsection{Box size dependence}
\label{sec:BCSbox}

The BCS approximation becomes sensitive to the size of the numerical expansion basis when many continuum states come into play~\cite{Migdal67}. 
This manifests in coordinate-space calculations primarily as a dependence on the numerical box size and grid spacing.
Since the continuum is artificially discretized due to a finite box size,  the density of the discretized states increases with increasing box size and effectively enhances the pairing strength logarithmically~\cite{Brack1972,Dobaczewski1996}. 
The effect is marginal in calculations that use a small pairing cutoff with well-bound nuclei, as the continuum states contribute little in that case. 
For larger pairing cutoffs and/or weakly bound nuclei, this effect grows and introduces a considerable numerical uncertainty. %instability.
In HFB calculations, however, the continuum states are localized, resulting in a weaker dependence on the size of the box.

\subsection{Localization of HFB canonical states}
\label{sec:HFB-1D2D}
    
The set of canonical single-particle (s.p.) wave functions $\varphi_\alpha$ is determined in the canonical representation by solving a system of coupled mean-field equations and taking the orthonormality of canonical states into account. The constrained variation of the total energy with respect to $\varphi_\alpha^\dagger$ results in the HFB
mean-field Hamiltonian~\cite{Dobaczewski1996,Reinhard1997, Tajima2004,Chen2022}
\begin{subequations}
\begin{equation}
  \hat{h}_{\mathrm{HFB},\alpha}
  =
  \hat{h}_\mathrm{mf}
  +
  \frac{u_\alpha}{v_\alpha}\hat{\tilde{h}}(\vec{r}),
\label{eq:HFB-Ham}
\end{equation}
where $\hat{h}_\mathrm{mf}$ is the standard mean-field
Hamiltonian, $\hat{\tilde{h}}(\vec{r})$ is the pairing potential, and $u_\alpha$,
$v_\alpha$ are the canonical occupation amplitudes. Note that, similar to Ref.~\cite{Chen2022}, the pairing-rearrangement term is treated as part of the mean-field Hamiltonian. The canonical s.p. energy and the canonical s.p. pairing gap are: 
\begin{equation}
  \varepsilon_\alpha=\langle\varphi_\alpha|\hat{h}_\mathrm{mf}|\varphi_\alpha\rangle
  \quad\mathrm{and}\quad
  \Delta_{\alpha\alpha}=\left|\langle\varphi_\alpha|\hat{\tilde{h}}(\vec{r})|\varphi_\alpha\rangle\right|,
  \quad
\label{eq:spenerg}
\end{equation} 
respectively.
The gap equation
\begin{equation} \label{eq:Evariation}
    0 = 4v_{\alpha}(\varepsilon_{\alpha} - \varepsilon_F) + 2\left(\frac{v_{\alpha}^2}{u_{\alpha}} - u_{\alpha}\right)\Delta_{\alpha\alpha},
\end{equation}
where $\varepsilon_F$ is the Lagrange multiplier for the particle number constraint (Fermi energy), is solved to get

% The solutions to the gap equation are then:
\begin{equation}
  \begin{Bmatrix}
       v_\alpha \\ u_\alpha
  \end{Bmatrix}
  =
\sqrt{\frac{1}{2}\left(1\mp\frac{\varepsilon_{\alpha}-\varepsilon_\mathrm{F}}{\sqrt{(\varepsilon_{\alpha}-\varepsilon_\mathrm{F})^2+\Delta_{\alpha\alpha}^2}}\right)}.
\end{equation} 
\end{subequations}

As the canonical continuum states are weakly occupied, i.e, $u_\alpha/v_\alpha\gg 1$, the  pairing potential in Eq.~(\ref{eq:HFB-Ham}) dominates, leading to localization given that $\hat{\tilde{h}}(\vec{r}) \propto \breve{\rho}(\vec{r})$
% \begin{equation}
%     \hat{\tilde{h}}(\vec{r}) \propto \breve{\rho}(\vec{r})
% \end{equation} 
and the pairing density $\breve{\rho}$ is confined to the nucleus, mostly at the nuclear surface. This means that the pairing correlations are similarly confined. Far from the nucleus, pairing density vanishes, and continuum quasiparticles behave as free particles. Consequently, HFB calculations are less sensitive to the chosen box size~\cite{Dobaczewski1996,Grasso2001,Grasso2002,Pei2011, Zhang2013}.

This fact renders results practically independent of the size of the box, but creates a new
problem: the pairing localization changes with the dimensionality and geometry of the implementation. In a 1D representation, the pairing potential is spherical and can localize only
in the radial direction, thus representing a potential well on the surface
of a sphere. In a 2D  representation, the imposed axial
symmetry causes the pairing potential to localize along the symmetry axis, which results in less spatial averaging and a deeper pairing potential well. In a Cartesian 3D code, even stronger localization is possible~\cite{Chen2022} given the unrestricted spatial symmetry. 

\label{sec:3Dpairing}
\begin{figure}
\centerline{\includegraphics[width=\linewidth]{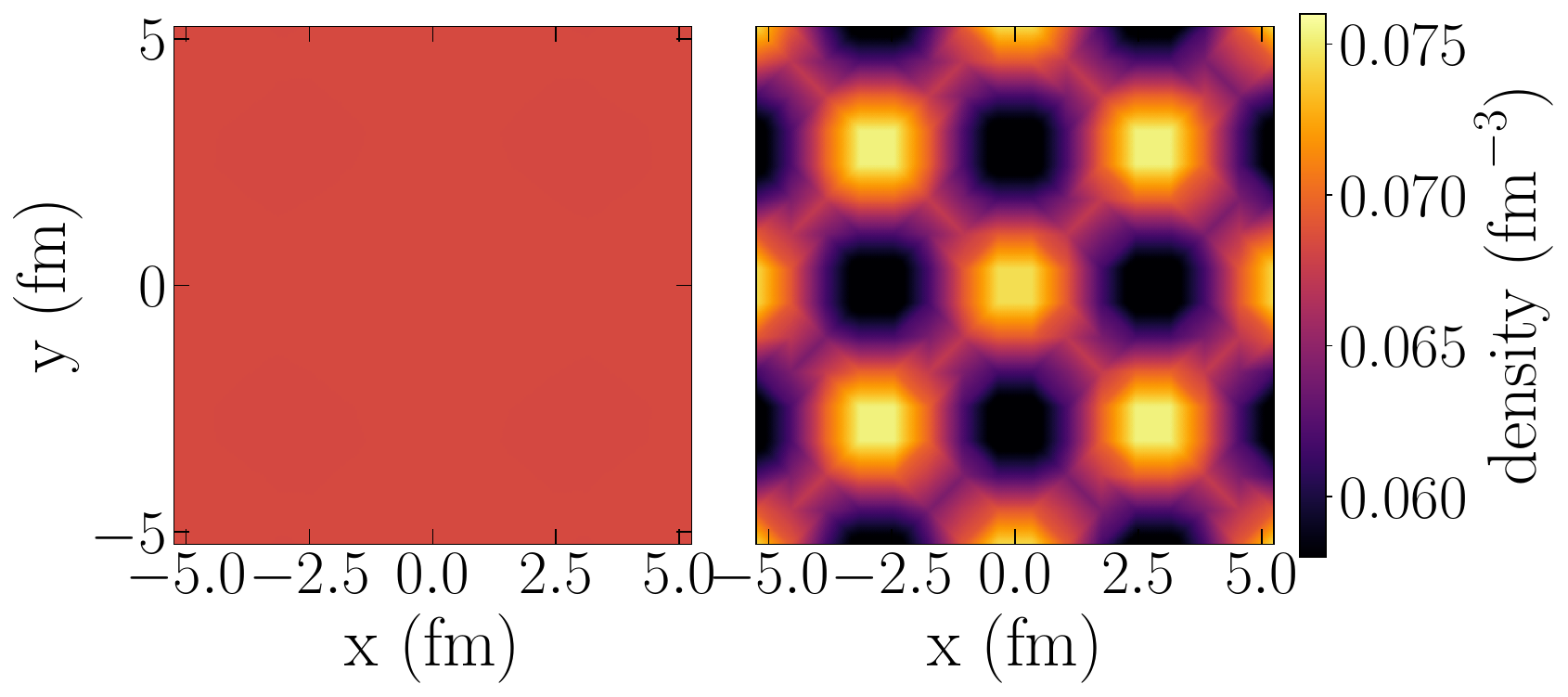}}
\caption{Cut of the density distributions along $x$- and $y$-plane of
  a Fermi gas with density-dependent zero-range pairing simulated in a
  numerical box in three dimensions with box length 11.2 fm and with
  average density 0.07/fm$^3$. The pairing parameters were taken from
  typical nuclear functionals and are $V_\mathrm{prot}=601$
  MeV/fm$^3$, $V_\mathrm{neut}=567$ MeV/fm$^3$, and switching density
  $\rho_{0,\mathrm{pair}}=0.212/\mathrm{fm}^3$. The left panel shows
  the BCS result and the right panel the HFB result.}
\label{fig:HFB-localization}
\end{figure}
A particular example of HFB localization can be seen for the case of infinite
matter. To demonstrate this, we have simulated a Fermi gas with a
density-dependent zero-range pairing functional
\begin{equation*}
  E_\mathrm{pair} =
  \sum_{\mathrm{nucl}\in\{\mathrm{p},\mathrm{n}\}}\frac{V_\mathrm{nucl}}{2}\int d^3r
  \left(1-\frac{\rho(\vec{r})}{\rho_{0,\mathrm{pair}}}\right)
  \breve{\rho}^*(\vec{r})\breve{\rho}(\vec{r})
  \;.
\end{equation*}
The resulting density distributions for BCS and HFB are shown in
Fig.~\ref{fig:HFB-localization}. BCS produces a perfectly homogeneous
density distribution while HFB leads to a strong inhomogeneity. It shows
up as a cubic lattice due to the cube-box boundary conditions. In infinite matter we expect that HFB will also break translational symmetry
resulting in some crystalline structure. Experience from plasma
physics suggests that the actual crystal symmetry depends on the
density~\cite{Mitchell1999, Piel2003}.
% The example shows that HFBis always prone to symmetry breaking due to localization.

The impact of dimensionality can be illustrated on the density of continuum canonical states, defined as
\begin{equation}
    g(\varepsilon) = \sum_k\delta(\varepsilon - \varepsilon_k),
\end{equation}
where $\varepsilon_k$ is the energy of the $k^{\mathrm{th}}$ s.p. state.
In practice, this is smoothed by a Gaussian to produce a continuous level-density function
\begin{equation}
    \tilde{g}({\varepsilon}) \approx \sum_k \frac{1}{\sqrt{\pi}\gamma}\exp\left(-\frac{(\varepsilon-\varepsilon_k)^2}{\gamma^2}\right), 
\end{equation}
where $\gamma$ is the smoothing width parameter. 
\begin{figure}
\centerline{\includegraphics[width=\linewidth]{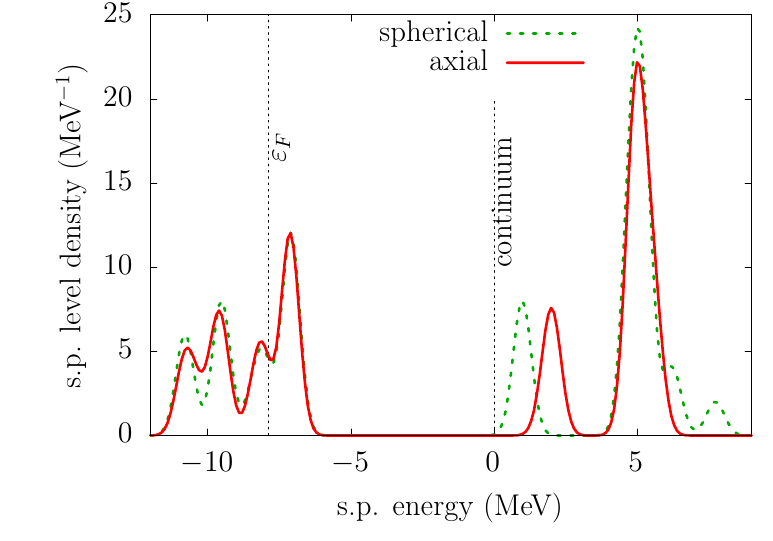}}
\caption{
Density of neutron s.p. states in $^{120}$Sn computed with HFB in spherical (dashed line) and axial (solid line) geometry using
the functional Fy(IVP). The green dashed line shows the 1D spherical result
and the red line the 2D axial result. The density of neutron s.p. states has been
smoothed by a Gaussian of 0.5 MeV width to render the graphical
representation better visible.
}
\label{fig:HFB_compare1D2D_120Sn}
\end{figure}
Figure~\ref{fig:HFB_compare1D2D_120Sn} shows the density of neutron s.p. states in $^{120}$Sn with respect to the s.p. energy using the 1D and 2D calculations. 
While the densities of bound states are similar, they show significant differences in the continuum region.
% Thus, localization changes with dimensionality of the implementation. 
%This demonstrates the change of localization with dimensionality. 
% One way to mitigate this dependence is via a soft pairing cutoff~\cite{Chen2022}.

A strong HFB localization creates another problem: pairing isomerism
from the state-wise pairing breakdown. This mechanism can be seen from the
HFB Hamiltonian (\ref{eq:HFB-Ham}). Small $v_\alpha$ localizes the
corresponding state, which, in turn, pushes the kinetic energy up and
with it the s.p. energy (\ref{eq:spenerg}).
A s.p. state may end up in a
local energy minimum with $v_\alpha=0$, leading to a pairing breakdown. 

\begin{figure}
\centerline{\includegraphics[width=\linewidth]{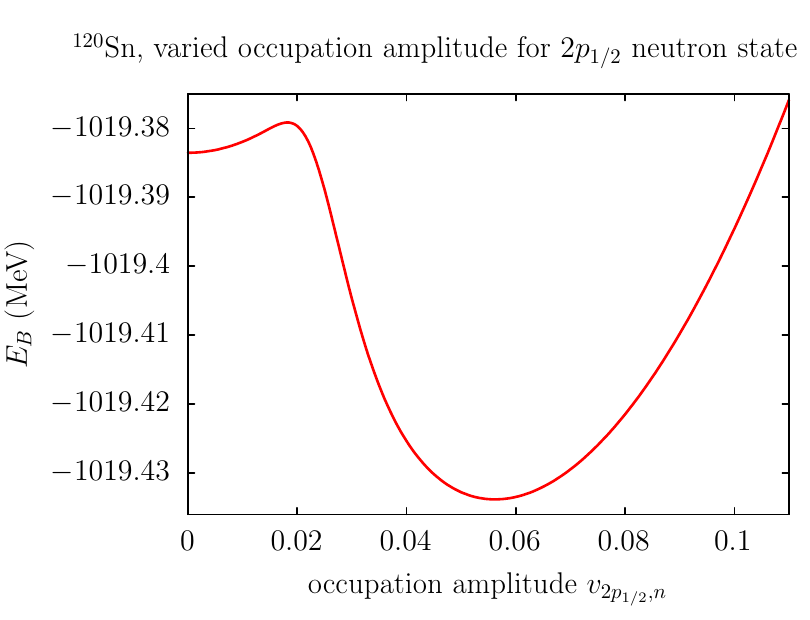}}
\caption{ Total energy  of  $^{120}$Sn computed with the functional SV-bas and HFB pairing as a function of constrained occupation amplitude
  $v_\alpha$ for the neutron $2p_{1/2}$ state which lies  in
  the continuum. }
\label{fig:HFB_vary_occ_120Sn_SVbas}
\end{figure}
To illustrate this, we perform an HFB calculation with one chosen
continuum state constrained to a given occupation $v_\alpha^2$ and
vary the constrained occupation systematically. The resulting energy curve is
shown in Fig. \ref{fig:HFB_vary_occ_120Sn_SVbas}.  It has two minima: the ground state with
finite $v_\alpha$ and the excited minimum with vanishing occupation in the neutron $2p_{1/2}$ state. This is only one of the many conceivable isomers in the continuum. This makes large-scale canonical-HFB
calculations  cumbersome because one has to check solutions
for possible isomerism. The expectation is that a softer localization, with localized s.p. wave functions that decay more slowly and have longer tails, 
keeps the kinetic energy lower and helps avoid the state-wise pairing
breakdown.

\section{Pairing functional with finite range}
\label{sec:finrange}
\subsection{Formalism}\label{sec:formalfold}
To achieve an efficient computational scheme, we
propose a separable approach, formulated in terms of an energy
density functional (here: for time-reversal symmetric systems) with the pairing functional:
\begin{equation} \label{eq:finrange-functional}
  E_\mathrm{pair}
  =
  \frac{1}{2}\int d^3rV(\rho({\vec{r}}))
  \breve{\rho}_{\mathcal{F}}^*(\vec{r})\breve{\rho}_{\mathcal{F}}(\vec{r})
  \;,
\end{equation}
where the local pairing interaction $V(\rho(\vec{r}))$ can be  density-dependent,
including gradient-density terms. The pairing density $  \breve{\rho}_{\mathcal{F}}(\vec{r})$
in (\ref{eq:finrange-functional})
is given by:
\begin{equation} \label{eq:finrange-functional1}
  \breve{\rho}_{\mathcal{F}}(\vec{r})
  =
  \sum_{\alpha>0} u_\alpha v_\alpha
  \left(\hat{\mathcal{F}}\varphi_\alpha\right)^\dagger(\vec{r})
  \left(\hat{\mathcal{F}}\varphi_\alpha\right)(\vec{r}),
\end{equation} 
where
the folding Hermitian operator $\hat{\mathcal{F}}$ can either be given in  an integral form:
\begin{equation}
  \left(\hat{\mathcal{F}}\varphi_\alpha\right)(\vec{r})
  =
  \int d^3r'F(\vec{r}-\vec{r}')\varphi_\alpha(\vec{r}'),
\end{equation}
where $F$ is the folding function, or as a differential operator associated with the momentum 
operator $\hat{\vec{p}}$:
\begin{equation}
  \left(\hat{\mathcal{F}}\varphi_\alpha\right)(\vec{r})
  =
  \tilde{F}(\hat{\vec{p}})\varphi_\alpha(\vec{r})
\end{equation}
with
\begin{equation}
  \tilde{F}(\vec{p})
  =
  \int d^3r\,e^{\mathrm{i}\vec{p}\cdot\vec{r}}\mathcal{F}(\vec{r}).
\end{equation}
Here and in the following, we adopt natural units  with $\hbar=1$.

The variation with respect to $u_\alpha, v_\alpha$ together with the
hermiticity of  $\hat{\mathcal{F}}$ yield the folded pairing potential:
\begin{equation}
\hat{\tilde{h}}_{\cal F}(\vec{r})=
  \hat{\mathcal{F}}V(\vec{r})\breve{\rho}_{\mathcal{F}}(\vec{r})\hat{\mathcal{F}}
\end{equation}
and the state-dependent pairing gap
\begin{equation}
\Delta_\alpha
  =
  \int d^3r\varphi_\alpha^\dagger
  \hat{\tilde{h}}_{\cal F}(\vec{r})
  \varphi_\alpha\, .
\end{equation}
 The state-dependent s.p. HFB Hamiltonian now becomes
\begin{equation}
  \hat{h}_{\mathrm{HFB},\alpha}
  =
  \hat{h}_\mathrm{mf}
  +
  \frac{u_\alpha}{v_\alpha}\hat{\tilde{h}}_{\cal F}(\vec{r}),
\end{equation}
where the pairing potential $\hat{\tilde{h}}(\vec{r})$ in Eq.~(\ref{eq:HFB-Ham}) has been replaced
by $\hat{\tilde{h}}_{\cal F}(\vec{r})$.

The pairing energy (\ref{eq:finrange-functional}) employs
the folded pairing density $\breve{\rho}_{\mathcal{F}}$  that contains two convoluted wave functions. This means that the present finite-range ansatz employs a fourfold separability.  The advantage of this prescription is that it
combines momentum conservation with the simple expression
(\ref{eq:finrange-functional}) for the pairing density functional. The twofold separable ansatz of Ref.~\cite{Tian2009} requires an
additional $\delta$-function for momentum conservation, which, in turn, inhibits a simple reduction to an energy-density functional.

%Moreover, the folding applied directly to the s.p. waevfunctions means that the local pairing potential $\hat{\tilde{h}}_{\cal F}(\vec{r})$ at $\vec{r}$ depends on the wavefunctions in a finite vicintiy of $\vec{r}$. This means that the finite-range pairing functional (\ref{eq:finrange-functional}) is non-local.

For the folding prescription, we adopt the
Gaussian folding:
\begin{equation}
  \mathcal{F}(\vec{r})
  \propto
  \exp\left(-\frac{\vec{r}^2}{\tilde{R}_{\mathcal{F}}^2}\right)
  \quad\longleftrightarrow\quad
  \tilde{F}\propto\exp\left(-\frac{\tilde{R}_{\mathcal{F}}^2}{4}\hat{\vec{p}}^2\right).
\label{eq:Gauss}
\end{equation}
Gaussian folding functions offer several advantages.  In grid representations, they allow for an efficient implementation by factorizing the Gaussians in any orthogonal coordinate system. While not employed in the current work, the Gaussian form is also well suited to the oscillator basis representation, making it beneficial to many HFB production codes. 
% Another advantage is the fast convergence of Gaussians. 
However, it should be noted that Fourier transformations (backwards and forwards) can get expensive depending on grid size. For the implementation on a 1D spherical grid, we apply folding as a matrix operation (see Appendix \ref{seq:matrix1D} for details).

\begin{figure}[bht]
\centerline{\includegraphics[width=\linewidth]{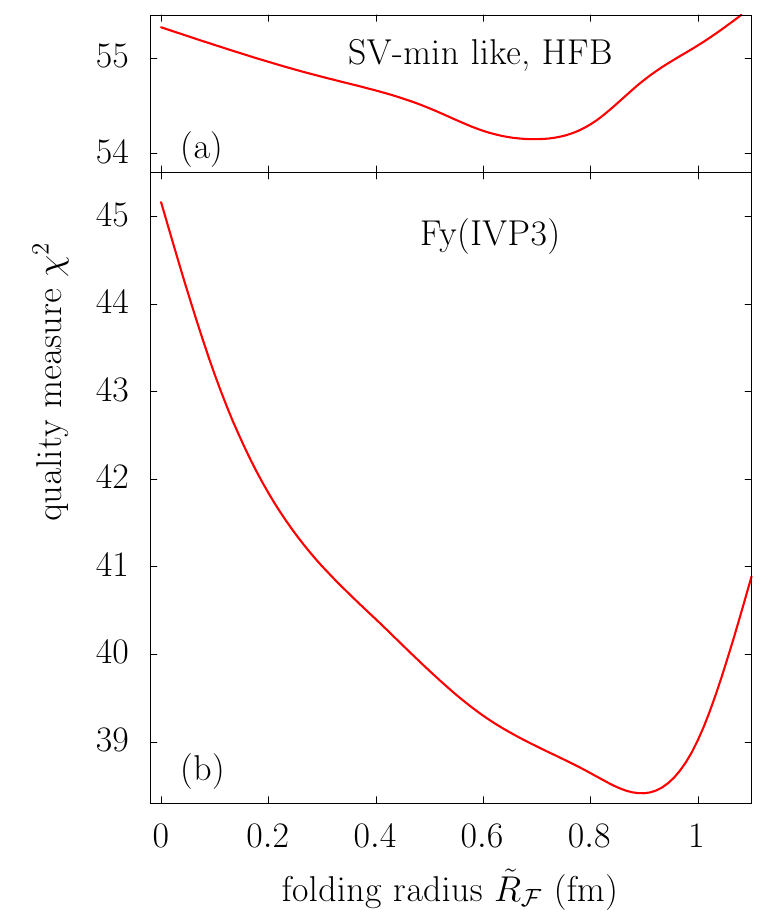}}
\caption{
The global quality measure $\chi^2$ as a function of the folding radius for two functionals: (a) Skyrme EDF   and  (b) Fayans EDF Fy(IVP3). Both EDFs were optimized with a cutoff in pairing space of $E_\mathrm{cut}=15$\,MeV. 
}
\label{fig:fit-width}
\end{figure}

\subsection{Optimization of functionals}
\label{sec:optim}

Using the prescription given in
Sec.~\ref{sec:formalfold}, we now attempt to quantify how the finite HFB folding radius affects the quality of a functional. To check this, we have performed $\chi^2$ fits with systematically varied folding radius. The calibration was done exactly in the same manner as for the corresponding zero-range functionals. For the Skyrme  functional with finite-range pairing, we  used the calibration dataset of SV-min, which consists of binding energies, r.m.s. charge radii, box-equivalent radii, and surface thicknesses for a selection of semi-magic nuclei from $^{16}$O up to $^{218}$U complemented by pairing gaps and spin-orbit splittings in doubly-magic nuclei. For more details, see Ref.~\cite{Kluepfel2009}. For Fayans functionals, we used the same data set as Fy(IVP), which consists of the  SV-min dataset with additional information on radius differences in Ca, Sn, and Pb isotopes\cite{Miller2019,Reinhard2024,Karthein2024}. For both functionals, the fits were done at the HFB level with a soft pairing cutoff of 15 MeV. The resulting quality measure $\chi^2$ is shown in Fig.~\ref{fig:fit-width}. The patterns are similar for both functionals: there is a preference for moderate folding radii between 0.7 and 0.9\,fm.
In Appendix~\ref{sec:modelparams} we provide the resulting functional   parameters at the $\chi^2$ minimum for both cases: $\tilde{R}_{\mathcal{F}} = 0.7$ fm for the Skyrme functional and 0.9 fm for the Fayans functional. The variation is rather small for the Skyrme calibration (upper panel), but 
Fy(IVP3) benefits substantially from a finite folding radius. 

\section{Results}
\label{sec:results}
% In this section, we explore the effects of folding on the properties of pairing functionals as applied to spherical and deformed nuclei.  We then study the portability of different numerical representations. Finally, we discuss optimization of finite-range functionals. 

For our tests, we choose $^{120}$Sn, which has a closed proton
shell (thus no proton pairing) and an open neutron shell. We use the SV-bas mean-field parametrization~\cite{Kluepfel2009} as a baseline for the EDF. In Appendix~\ref{sec:DFT} we provide the details of the functionals used in this work. Note that the published neutron pairing strength applies only to a soft cutoff with the scheme as specified in Ref.~\cite{Kluepfel2009}. In our case, the neutron pairing strength changes dramatically with pairing scheme, cutoff, and folding radius. Thus, we must adjust it to the given folding parameters and cutoff energies, as an isolated change in model parameters (here, the folding radius) can compromise the quality of the fit.  For example, if the neutron pairing strength for folding radius $\tilde{R}_{\mathcal{F}}=0$ is 317 MeV$\,$fm$^2$, it becomes 4340 MeV$\,$fm$^2$ for $\tilde{R}_{\mathcal{F}}=1.5$\,fm and 26700 MeV$\,$fm$^2$ for $\tilde{R}_{\mathcal{F}} = 3$\,fm.  Unless otherwise specified, the other parameters of the functional used in this section were kept at the values optimized at $\tilde{R}_{\mathcal{F}} = 0$\,fm.

\begin{figure*}[htbp]
\centerline{\includegraphics[width=0.9\linewidth]{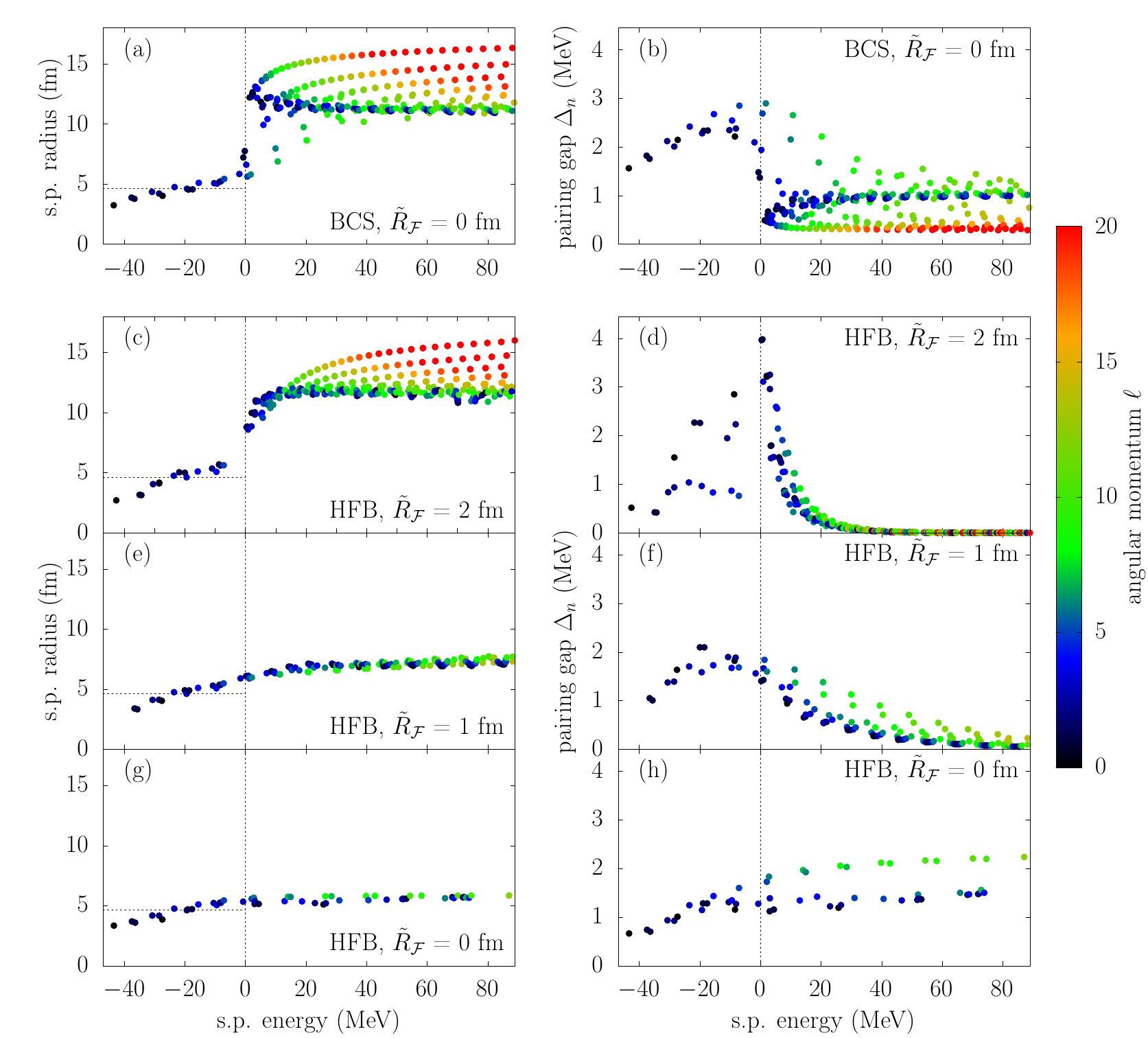}}
\caption{Trends of key observables for BCS with $\tilde{R}_{\mathcal{F}} = 0$ (panels (a) and (b)) and HFB (panels (c)-(h)) for three different folding radii, $\tilde{R}_{\mathcal{F}}$ = 0, 1, and  2 fm as functions of s.p. energy. Test case is $^{120}$Sn computed with SV-bas and cutoff energy $\varepsilon_\mathrm{cut}=100$ MeV. The orbital angular momentum $\ell$ of the s.p. states is indicated by color code. Left panels: s.p. r.m.s. radii are plotted as a function of s.p. energy. The radius scale extends up to 18 fm, corresponding to the radius of the numerical box. Right panels: neutron s.p. pairing gaps $\Delta_\alpha$ plotted as a function of s.p. energy.
The faint vertical lines indicate the continuum threshold. The  horizontal dotted line indicates the total charge r.m.s. radius of $^{120}$Sn.}
\label{fig:varyocc_widths}
\end{figure*}

\subsection{Impact of folding range on observables}
\label{sec:range-sp}

Figure~\ref{fig:varyocc_widths} shows trends of observables for four different cases: BCS at $\tilde{R}_{\mathcal{F}} = 0$ and HFB at $\tilde{R}_{\mathcal{F}} = 0, 1,$ and $2$\,fm. 
The left panels show the distribution of s.p. radii as a function of the s.p. energy. The colors of the points indicate the orbital angular momentum $\ell$ of a state. For the BCS case with the zero folding radius, the s.p. radii show a marked change at the continuum threshold. Below, they grow steadily with increasing s.p. energy, while at the threshold, they jump to large values. The continuum states with radii around 15\,fm extend over the numerical box, while a few states have smaller radii. These are weakly localized continuum resonances.

The HFB calculations produce a different trend. At zero folding, all s.p. radii are below 6\,fm. This is due to strong localization produced by the $u/v$ contribution in the HFB Hamiltonian (Eq. \ref{eq:HFB-Ham})~\cite{Dobaczewski1984, Dobaczewski1996, Miller2019}. Increasing folding radius, weakens the localization until it is practically lost for $\tilde{R}_\mathcal{F}=2$ fm. With few exceptions, 
the s.p. radii of the unbound states  increase with $\ell$, which is an effect of the centrifugal term. It is interesting to note that HFB localization, which is strong for $\tilde{R}_\mathcal{F}=0$ and still appreciable for $\tilde{R}_\mathcal{F}=1$ fm, pushes the high-$\ell$ states to higher energies beyond those shown.

The right panels of Fig.~\ref{fig:varyocc_widths} show the distribution of s.p. neutron pairing gaps $\Delta_\alpha$. The pairing gaps generally increase  up to the zero-energy  continuum threshold. Above the threshold, most of the BCS gaps 
drop quickly down to very small values, with  a few remaining at moderate values for the localized continuum resonances. 
The trends in gaps  also show a marked change with the folding radius in
relation to the changes in radii. In the HFB variant at $\tilde{R}_{\mathcal{F}} = 0$,  the continuum states have  large
gaps because all these
 states are well localized and the interaction is independent of s.p. energy. 
 A nonzero finite folding radius leads to a decrease in
the interaction with increasing kinetic energy.   This is  seen for the
two cases with non-zero folding radii. For $\tilde{R}_{\mathcal{F}}=1$\,fm,  pairing gaps vary smoothly across the threshold, while for $\tilde{R}_{\mathcal{F}}=2$\,fm,  there appears a sharp threshold peak.

The results shown in Fig.~\ref{fig:varyocc_widths} suggest that large folding radii are disadvantageous. On the other hand, folding radii around $\tilde{R}_{\mathcal{F}}=1$ fm promise a fair compromise between convergence, soft localization, and smooth trends in physical properties.

A word is in order about the HFB isomers discussed around
Fig.~\ref{fig:HFB_vary_occ_120Sn_SVbas} and often seen in connection with zero-range pairing. The BCS calculations are free of that problem.  However, in HFB calculations with zero folding radius, pairing isomers are likely to develop 
(see Fig.~\ref{fig:HFB_vary_occ_120Sn_SVbas}). This is not the case for larger folding radii: in our calculations, we did not find HFB isomers for $\tilde{R}_{\mathcal{F}} \ge 0.5$~fm.

The cutoff energy $\varepsilon_\mathrm{cut}$ in pairing space controls the number of active  s.p. states $N_{\text{st}}$ to be included in the calculation~\cite{Reinhard2021}. In this section, we check the convergence of the results with increasing pairing space. We expect that a combination of the cutoff energy with a finite folding radius would improve the convergence.

\subsection{Dependence on the size of pairing space}
\label{sec:varyspace}

Even though finite-range pairing renders the summations over s.p. states convergent, they still may converge slowly, thus requiring very many  s.p. states. Practical considerations often suggest to limit their number. The first measure is technical; one limits the number of s.p. states in the calculations, either by number or by maximum s.p. energy. However, this hard cutoff can lead to fluctuations in the results when changing the numerical representation, the nucleus, or its deformation. To avoid this, a soft cutoff is used in a second step, as described in Sec. \ref{sec:softcut}. Here, we investigate the impact of the number of active states on the results.

\begin{figure}
\centerline{\includegraphics[width=\linewidth]{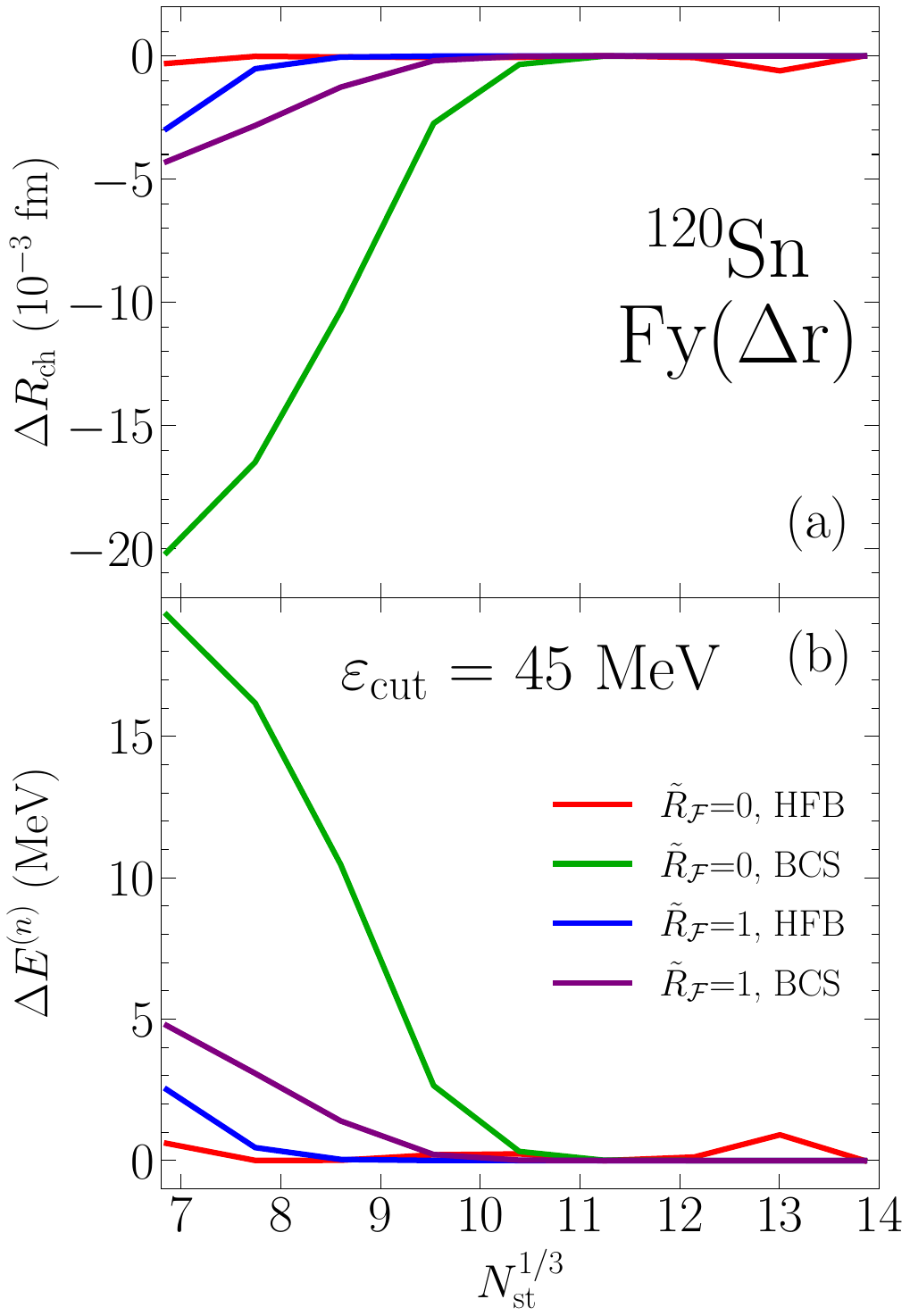}}
\caption{Relative (a) charge radius $(\Delta R_{\rm{ch}} = R_{\rm{ch}} - R_{\rm{ch,\,ref}})$  and (b) neutron pairing energy  $\left(\Delta E^{(n)} = E^{(n)} - E_{\rm{ref}}^{(n)}\right)$ 
in $^{120}$Sn versus size of pairing space  $N^{1/3}_\mathrm{st}$  computed with Fy($\Delta r$) EDF~\cite{Reinhard2017} with
$\varepsilon_\mathrm{cut}=45$\,MeV. $R_{\rm{ch,\,ref}}$ and $E_{\rm{ref}}^{(n)}$ are reference values obtained at the largest pairing space.
%To visualize the variation, results are shown relative to the values obtained at the largest pairing space.
}
\label{fig:Sn120-vardeocc}
\end{figure}

Figure~\ref{fig:Sn120-vardeocc} shows the dependence of the charge radius and pairing energy in the spherical $^{120}$Sn on the cube root of the number of states  $N_\mathrm{st}$ for $\epsilon_{\mathrm{cut}} = 45$\,MeV. The deviations  are significant  at the smallest pairing spaces. 
The worst case is that of BCS at zero folding radius. At $\tilde{R}_{\mathcal{F}} = 1$~fm, the fluctuations are reduced  by an order of magnitude.  
%With a comfortably small number of $N_{\mathrm{shells}}$, one already achieves reliable results. 
%However, the case of $\tilde{R}_{\mathcal{F}} = 1$~fm deviates more at $\Delta_\mathrm{shell}=2$, although it is unimportant in practice as one uses 3 extra shells to be on the safe side.

%\subsection{A deformed nucleus example}

\begin{figure}[htbp]
\centerline{\includegraphics[width=\linewidth]{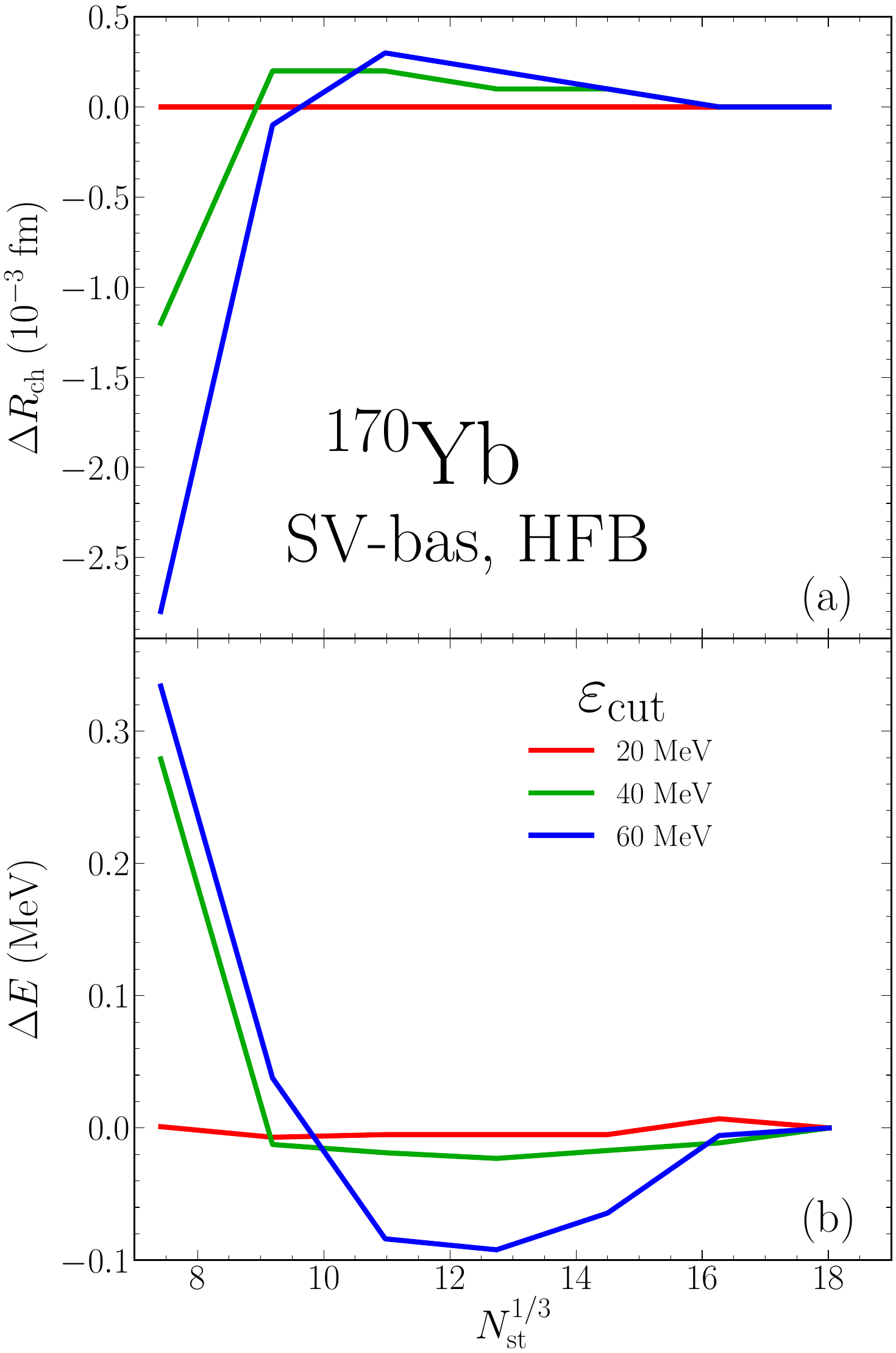}}
\caption{Relative (a) charge radius $(\Delta R_{\rm{ch}} = R_{\rm{ch}} - R_{\rm{ch,\,ref}})$
	and (b) total energy $(\Delta E = E - E_{\rm{ref}})$ versus size of pairing space  $N^{1/3}_\mathrm{st}$ of the deformed nucleus $^{170}$Yb for different cutoff energies computed with
	SV-bas and HFB. $R_{\rm{ch,\,ref}}$ and $E_{\rm{ref}}$ are reference values obtained at the largest pairing space. The folding radius is  $\tilde{R}_{\mathcal{F}} = 1$\,fm.}
\label{fig:trend-deocc-Yb170}
\end{figure}

In addition, we have also investigated a case of deformed $^{170}$Yb using the \texttt{SkyAx} code~\cite{Reinhard2021}, which employs a cylindrical coordinate-space grid. While the convergence of results in this case is slower, we find in all respects the same behavior as for spherical nuclei. The dependence of total binding energy and charge radius on the size of pairing space in terms of $N_\text{st}$ is demonstrated in Fig.~\ref{fig:trend-deocc-Yb170}. In these calculations, the folding radius was set to $\tilde{R}_\mathcal{F} = 1$~fm.  The convergence is fast for cutoff energies up to 40\,MeV.   
The largest cutoff of\,60 MeV shows more $N_\mathrm{st}$-dependence. However, one should keep in mind the scale: in the space above $N_{\mathrm{st}}^{1/3} = 9$, the resulting error on the radius
remains below 0.0005\,fm and it is below 0.1\,MeV for the total energy; both are acceptable uncertainties for most practical applications.

\subsection{Cutoff dependence}
\label{sec:range-cutoff}

Having answered the  question of the  number of states in the s.p. basis, we now discuss the impact of the cutoff energy. In all cases, we ensure that the number of states is always large enough to cover the given cutoff energy.

\begin{figure}[htp]
\centerline{\includegraphics[width=\linewidth]{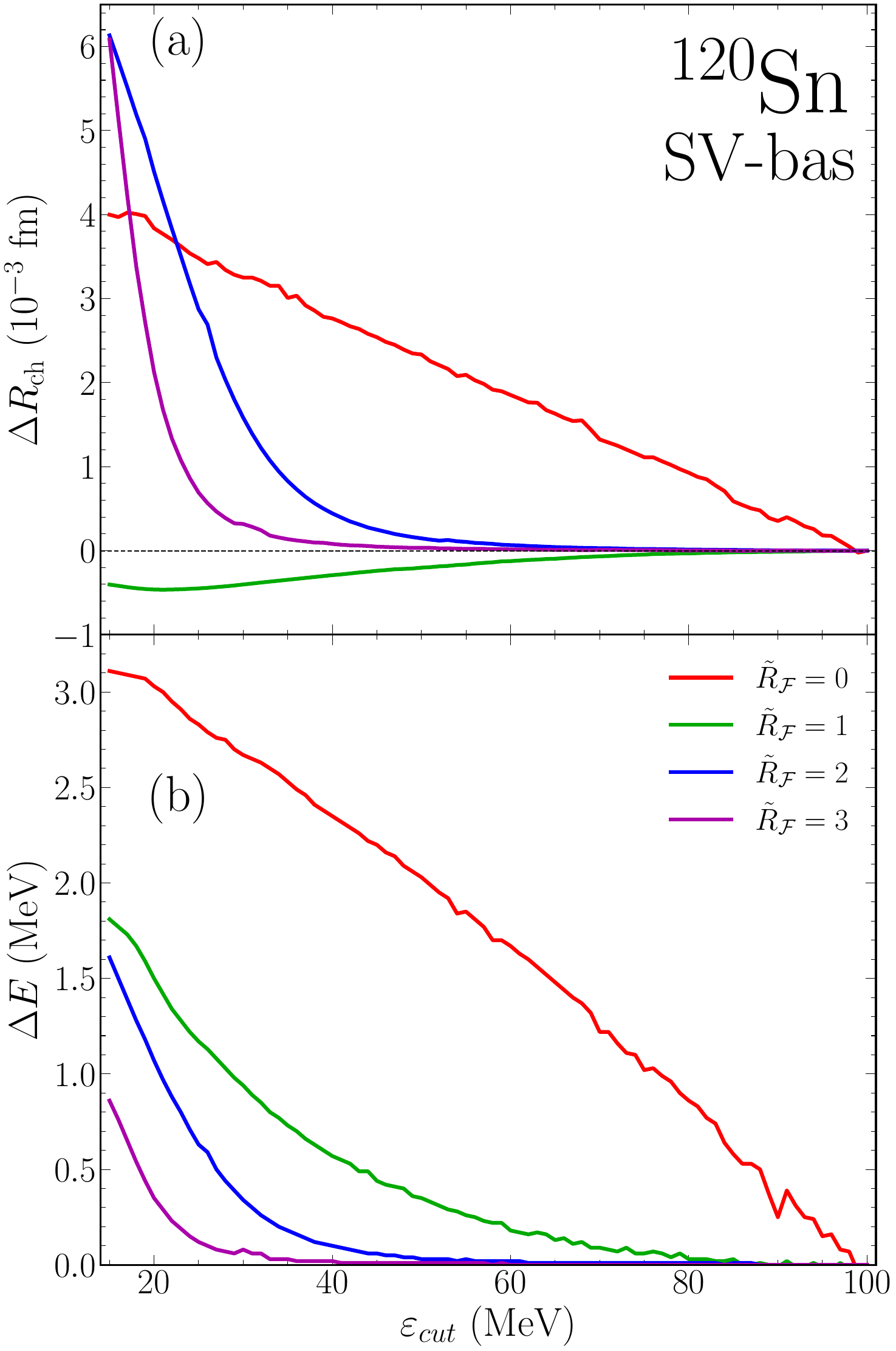}}
\caption{Relative (a) charge radius  $(\Delta R_{\rm{ch}} = R_{\rm{ch}} - R_{\rm{ch,\,ref}})$
	and (b) total energy $(\Delta E = E - E_{\rm{ref}})$ as a function of the cutoff energy
  $\varepsilon_\mathrm{cut}$ for $^{120}$Sn computed with SV-bas and HFB
  pairing for four different folding radii $\tilde{R}_{\mathcal{F}}$ as indicated (in units of fm). The neutron pairing
strength has been calibrated for each folding radius such that the pairing gap
is the same at $\varepsilon_\mathrm{cut}=100$\,MeV. $R_{\rm{ch,\,ref}}$ and $E_{\rm{ref}} $ are the reference values of the charge radius and total energy, respectively, at  $\varepsilon_\mathrm{cut}=100$\,MeV.
}
%To emphasize the relative changes, all observables are scaled with respect to the reference values, $R_{\rm{ch,\,ref}}$ and $E_{\rm{ref}} $,  at
% $\varepsilon_\mathrm{cut}=100$\,MeV.
\label{fig:varycut-120Sn_DiffHFB}
\end{figure}

To illustrate the effect of the varying cutoff  energy,
Fig.~\ref{fig:varycut-120Sn_DiffHFB} shows the charge radius and total HFB energy  at different folding radii as a function of $\varepsilon_\mathrm{cut}$. To better illustrate the variation with cutoff, all observables are plotted relative to their value at  $\varepsilon_\mathrm{cut}=100$\,MeV. The pairing strength has been calibrated for each folding radius to deliver the same pairing gap at
$\varepsilon_\mathrm{cut}=100$ MeV. The observables converge nicely with the increasing cutoff for the nonzero folding radii. An increase in the folding radius results in a faster convergence. No convergence is seen for $\tilde{R}_{\mathcal{F}} = 0$, since pairing properties are known to diverge for zero-range pairing interactions~\cite{Bulgac2001a,Bulgac2002}.

Although finite-range pairing creates an ultimately converging pairing scheme, the actual rate of convergence is moderate. It is only at  large folding radii that  one can achieve near full convergence at  cutoff energies of about 30 MeV. 
However, such large folding radii are excluded in calculations due to quality considerations (see Sec. \ref{sec:optim}). 

\subsection{Box-size  dependence}
\label{sec:boxdep}

\begin{figure}[htpb]
\centerline{\includegraphics[width=\linewidth]{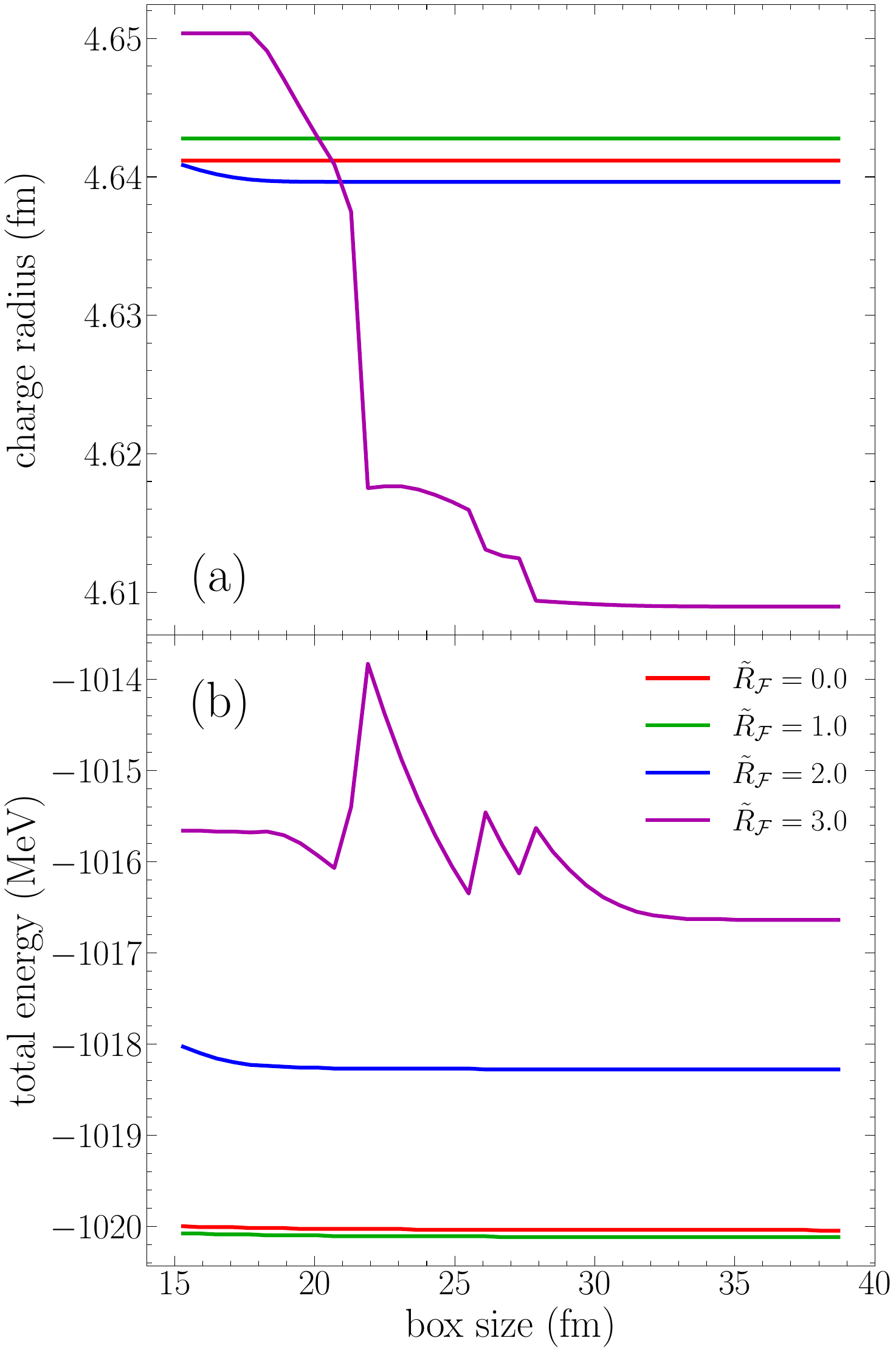}}
\caption{(a) Charge radius and (b) total HFB energy versus box size computed at $\varepsilon_\mathrm{cut}=25$ MeV with $\tilde{R}_{\mathcal{F}} = 0, 1, 2$, and  3\,fm for the case of $^{120}$Sn using the SV-bas functional. The kinks for $\tilde{R}_{\mathcal{F}} = 3$ are caused by unlocalized  continuum states.}
\label{fig:varycut-120Sn_hfb_box}
\end{figure}
In Sec.~\ref{sec:BCSbox}, we discussed the dependence of box size for BCS and HFB calculations with zero-range pairing. Here, we examine the changes associated with finite-range pairing. We concentrate on HFB and use a somewhat larger cutoff energy ($\varepsilon_{\text{cut}} = 25$ MeV) to enhance the impact of continuum states. Figure~\ref{fig:varycut-120Sn_hfb_box} shows trends of two key observables, namely, the total binding energy, and the charge radius as a function of box size for various folding radii as indicated, using the SV-bas functional. All HFB results show occasional cusps. This indicates that there is still a problem with HFB iterations being caught in a state-wise pairing isomerism. But the cusps are so small that the effects on the observables are insignificant.  
The case of $\tilde{R}_{\mathcal{F}}=1.0$~fm is as robust as the zero folding radius, though the situation worsens for larger radii. For a folding radius of 3~fm, we see a large box dependence in HFB for box radius $\lesssim$ 33 fm. This variation happens since the folding radius being too large reduces contribution from the continuum states, resulting in a numerical instability, similar to the BCS case. A large box size $\gtrsim$ 33 fm accounts for the continuum states resulting in converged solutions.

\subsection{Portability}
\label{sec:portability}

\begin{figure}[htp]
\centerline{\includegraphics[width=\linewidth]{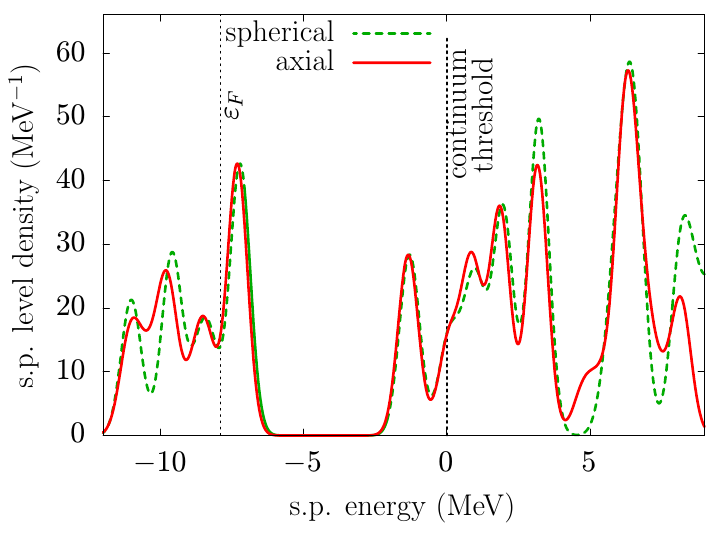}}
\caption{
Density of neutron s.p. states in $^{120}$Sn computed with HFB in spherical (dashed line) and axial (solid line) geometry using
the functional SV-bas and a folding radius $\tilde{R}_{\mathcal{F}} = 1.5$ fm. 
The density of states has been
smoothed by a Gaussian of 0.5 MeV width to render graphical representation better visible.
}
\label{fig:HFB_compare1D2D_120Sn_width15}
\end{figure}

Another argument for introducing a finite folding radius in the pairing functional is that it helps to improve the portability between different numerical representations.
In Fig.~\ref{fig:HFB_compare1D2D_120Sn}, we demonstrated that a zero-range pairing functional can lead to different spectral distributions near continuum threshold. 
Fig.~\ref{fig:HFB_compare1D2D_120Sn_width15} applies the same analysis
to a pairing functional with finite folding radius $\tilde{R}_{\mathcal{F}}=1.5$ fm. We see now a near-perfect agreement in the continuum region. %The slightly different height of the peaks is not a matter of concern. 
The axial peak heights are slightly lower and their widths slightly broader because the strictly degenerated s.p.~levels in the spherically symmetric grid are spread in an axial representation due to symmetry breaking. 
For the same reason, around -10 MeV, another broadening and slight shift is seen. 
%This could be improved by using large biox size and higher spatial resolution, however, at the price of considerably more numerical expense. We better leave figure \ref{fig:HFB_compare1D2D_120Sn_width15} as it is. It shows the improvement as compared to figure \ref{fig:HFB_compare1D2D_120Sn} and at the same time the remaining (small) influence of numerical parameters of the grid representation.

\section{Conclusions}\label{Conc}
In this work, we studied a finite range pairing functional
based on the Gaussian  folding of the canonical HFB wave 
functions entering the pairing density.
The proposed formalism has been applied to two DFT models based on Skyrme and Fayans energy density functionals.
When compared to BCS or HFB calculations with zero-range pairing, our finite-range pairing prescription allows one to choose lower cutoffs and thus smaller model spaces. The approach also helps avoid pairing isomer solutions associated with vanishing pairing, resulting in better convergence of HFB iterations. The finite-size pairing also improves the portability of the implementation between different spatial geometries. 

We  explored the effects of the size of the pairing space on the convergence of   BCS and HFB calculations. The HFB calculations with non-zero folding radii converge well with increasing pairing space  for both spherical and deformed nuclei. A combination of smaller folding radii ($\sim$1\,fm) and cutoff energies ($\sim$15 MeV) is sufficient for an accurate and stable calculation.  For large folding radii, such as $\tilde{R}_{\mathcal{F}} = 3$\,fm, the calculations show a strong box size dependence due to unlocalized s.p. states in continuum, similar to the benchmark BCS calculations.

Lastly, we have re-optimized the functionals with the finite folding radius and demonstrated a slightly enhanced calibration performance for both Skyrme and Fayans EDFs.
The improved performance and stability of the new EDFs are promising for large-scale calculations of nuclear properties, and a systematic study exploring the performance on other observables is currently underway.

\section*{Acknowledgements}
Discussions with Ante Ravli\'{c} are gratefully appreciated.
This material is based upon work supported by the U.S. Department of Energy under Award Numbers DE-SC0013365 and DE-SC0023688 (Office of Science, Office of Nuclear Physics), and DE-SC0023175 (Office of Science, NUCLEI SciDAC-5 collaboration). P.-G.R. acknowledges computing resources from the computing center of the University Erlangen/Nürnberg (RRZE).

%\clearpage

\appendix

\section{The functionals in detail}
\label{sec:DFT}

In this Appendix, we provide the details for the energy-density
functionals used in this paper. 

\subsection{The  local densities}
\label{sec:dens}

Nuclear energy-density functionals are usually expressed in terms of a 
local densities and currents. As we consider here only ground states of even-even
nuclei, we can limit ourselves to the time-even densities: local density $\rho$,
kinetic-energy density $\tau$, spin-orbit density ${\bf J}$, and pair
density ${\breve\rho}$.  They read:
\begin{subequations}
\label{eq:basdens}
\begin{eqnarray}
  \rho_q(\vec{r})
  &=&
  \sum_{\alpha\in q}f_\alpha^{\mathrm{(cut)}} v_\alpha^2\,
  \big|\varphi_\alpha(\vec{r})\big|^2,
\\
  \tau_q(\vec{r})
  &=&
  \sum_{\alpha\in q}f_\alpha^{\mathrm{(cut)}} v_\alpha^2\,
  \big|\nabla\varphi_{\alpha}(\vec{r})\big|^2,
\\
  {\bf J}_q(\vec{r})
  &=&
  -\mathrm{i}
  \sum_{\alpha\in q}f_\alpha^{\mathrm{(cut)}}v_\alpha^2\,
  \varphi_{\alpha}^+(\vec{r})\vec{\nabla}\!\times\!\hat{\vec{\sigma}}
  \varphi_{\alpha}^{\mbox{}}(\vec{r}),
\\
  {\breve\rho_p}(\vec{r})
  &=&
  2\sum_{\alpha>0, \alpha\in q}f_\alpha^{\mathrm{(cut)}} u_\alpha v_\alpha
  \big|\varphi_\alpha(\vec{r})\big|^2,
\end{eqnarray}
\end{subequations}
where $q=p$ or $n$.
It is often useful to express densities in terms of  isoscalar ($T=0$) and isovector ($T=1$) components. For instance:
\begin{equation}
  \rho_{T=0}
  =
  \rho_{\rm p}+\rho_{\rm n},\quad
  \rho_{T=1}
  =
  \rho_{\rm p}-\rho_{\rm n}.
\end{equation}

\subsection{Soft cutoff in s.p. space}
\label{sec:softcut}

The pairing occupation amplitude $v_\alpha$ decreases slowly with
s.p. energy $\varepsilon_\alpha$. Thus, the summations in (\ref{eq:basdens}) do  not converge  if all s.p. states are considered. 
A possible way to remedy this problem is to apply a
soft cutoff weight depending on s.p. energy $\varepsilon_\alpha$~\cite{Krieger1990}:
\begin{equation}
  f_\alpha^{\mathrm{(cut)}}
  =
  \frac{1}{1+\exp{\left[\varepsilon_\alpha-(\epsilon_\mathrm{F}-\delta\epsilon)\right]/(\delta\epsilon/10)}},
\end{equation}
where $\delta\epsilon$ is the energy span of the pairing active zone
above the Fermi surface. 
$\delta\epsilon$ is a crucial cutoff parameter~\cite{Bender2000, Reinhard2021}. Older calculations used 5 MeV, which embraces almost exclusively the bound states. The larger the cutoff, the more continuum states come into play.

\subsection{The Skyrme energy functional}
\label{sec:Skyrme-form}

The Skyrme functional for time-even energy density can be written  as \cite{Bender2003}
% \begin{subequations}
% \begin{eqnarray}
\begin{equation}
\label{eq:Eskeven}
  E  =  \int d^3r\,\left\{
    \mathcal{E}_{\rm kin}
    +\mathcal{E}_{\mathrm{Sk}}
    +\mathcal{E}_{\mathrm{Sk,pair}}
  \right\}
  +
  E_{\rm Coul}
  +
  E_{\rm cm},
% \end{eqnarray}
\end{equation}
where
\begin{eqnarray}
  \mathcal{E}_{\rm kin}
  &=& 
  \frac{\hbar^2}{2m}\,\tau, 
\label{eq:ekindens}
\nonumber \\
  \mathcal{E}_{\mathrm{Sk},T}
& = &\sum_{T=0}^1\Big[ C_T^{\rho}\rho_{T}^{2}
      + C_T^{\Delta \rho} \, \rho_{T} \Delta \rho_{T}
\nonumber\\
&   & 
      + C_T^{\tau} \, \rho_{T} \tau_{T}
      + C_T^{J} J^2_{T} 
      + C_T^{\nabla J} \rho_{T} \, \nabla\!\cdot\!\vec{J}_{T}\Big],
\nonumber\\
  \mathcal{E}_{\mathrm{Sk,pair}}
  &=&
  \frac{1}{4} 
  \left[1 -\frac{\rho}{\rho_\mathrm{pair}}\right]
  \sum_{q\in\{p,n\}}V_{0,q}\, {\breve\rho}_q^2
  \quad,
\label{eq:ep2}\nonumber\\
  {E}_{\rm Coul}
  &=& 
  e^2 \frac{1}{2} \iint d^3r\,d^3r'
        \frac{\rho_{\rm p}(\vec{r})\rho_{\rm p}(\vec{r}')}
             {|\vec{r}-\vec{r}'|}
\nonumber\\
  &&
      - \frac{3}{4}e^2\left( \frac{3}{\pi} \right)^{1/3}
        \int d^3r\rho_{\rm p}^{4/3}(\vec{r})
  \quad,
\label{eq:coulfun}\nonumber\\
  E_{\rm cm}^{\mbox{}}
  &=&
  -\frac{1}{2mA}\langle\big(\hat{P}_\mathrm{cm}\big)^2\rangle
  \quad,\quad
  \hat{P}_\mathrm{cm}
  =
  \sum_i\hat{p}_i
  \ . \nonumber
\label{eq:cmfull}
\end{eqnarray}
% \end{subequations}
A density dependence in the Skyrme energy is included only in the
volume term $\propto\rho_T^2$. 

\subsection{The Fayans energy functional}
\label{sec:Fayans-form}

Alternatively, we also discuss results obtained with the Fayans EDF~\cite{Reinhard2017}. The functional used in this work has a form where each of the three terms in the pairing functional is split into separate proton and neutron terms, each with its own strength coefficient. We abbreviate that as Fy(IVP3) to denote isospin variable pairing in all 3 terms. The functional reads:
% \begin{subequations}
\label{eq:Fyfunc}
\begin{eqnarray}
  E
  &=&
  \int d^3r\,\left\{
    \mathcal{E}_{\rm kin}
  +\mathcal{E}_{\mathrm{Fy}}^\mathrm{v}
  +\mathcal{E}_{\mathrm{Fy}}^\mathrm{s}
  +\mathcal{E}_{\mathrm{Fy}}^\mathrm{ls}
  +\mathcal{E}_{\mathrm{Fy}}^\mathrm{pair}
  \right\}
\nonumber\\
  &&
  + E_{\rm Coul}
  + E_{\rm cm},
\end{eqnarray}
where
\begin{eqnarray}
\mathcal{E}_{\mathrm{Fy}}^\mathrm{v}
  &=&
  \sfrac{\varepsilon_F\rho_\mathrm{sat}}{3}
  \left[
  {a_+^\mathrm{v}}
  \frac{1\!-\!{h_{1+}^\mathrm{v}}x_0^{{\sigma}}}
       {1\!+\!{h_{2+}^\mathrm{v}}x_0^{{\sigma}}}x_0^2
  \!+\!
  {a_-^\mathrm{v}}
  \frac{1\!-\!{h_{1-}^\mathrm{v}}x_0}
       {1\!+\!{h_{2-}^\mathrm{v}}x_0}x_1^2
  \right],
\label{EFay-dens}
\nonumber\\
  \mathcal{E}_\mathrm{Fy}^\mathrm{s}
  &=& 
  \sfrac{\varepsilon_F\rho_\mathrm{sat}}{3}
  \frac{a_+^\mathrm{s}r_s^2(\vec{\nabla} x_0)^2}
       {1 %+h_{+}^\mathrm{s}x_0^\sigma
        +h_{\nabla}^\mathrm{s}r_s^2(\vec{\nabla} x_0)^2},
\label{EFay-grad} 
\nonumber\\
  \mathcal{E}_\mathrm{Fy}^\mathrm{ls}
  &=&
  \sfrac{4\varepsilon_F r_s^2}{3\rho_\mathrm{sat}}
  \left(\kappa\rho_0\vec{\nabla}\cdot\vec{J}_0
  +
  \kappa'\rho_1\vec{\nabla}\cdot\vec{J}_1  
%  +g \vec{J}_0^2 + g'\vec{J}_1^2
  \right),
\label{eq:EFy-ls}
\nonumber\\
  \mathcal{E}_{\mathrm{Fy}}^\mathrm{pair}
  &=&
  \sfrac{2\varepsilon_F}{3\rho_\mathrm{sat}}
    \sum_{t=p,n}{\breve\rho}_t^2
  \Big[f_{\mathrm{ex},t}^\xi
       +h_t^\xi x_\mathrm{pair}^{\gamma}
\nonumber\\
  && \qquad\qquad
 +h_{\nabla,t}^\xi r_s^2 (\vec{\nabla} x_\mathrm{pair})^2\Big]
 \nonumber
\label{eq:epF}
\end{eqnarray}
% \end{subequations}
with $x_T=\rho_T/\rho_{\mathrm{sat}}$. The Coulomb and
c.m. terms are the same as in eq.~(\ref{eq:Eskeven}). Several model
parameters are fixed a priori. These include the nucleon masses
$\hbar^2/2m_p=20.749811$\,MeV\,fm$^2$,
$\hbar^2/2m_n=20.721249$\,MeV\,fm$^2$, the charge
$e^2=1.43996448$\,MeV\,fm, the reference saturation density
$\rho_\mathrm{sat}=0.16$\,fm$^{-3}$,
$\rho_\mathrm{pair}=\rho_\mathrm{sat}$, $\sigma=1/3$, and
$\gamma=2/3$.  The saturation density $\rho_\mathrm{sat}$ also determines
the auxiliary parameters, namely, the Wigner-Seitz radius $r_s=
(3/(4\pi\rho_\mathrm{sat}))^{1/3}$\,fm and Fermi energy
$\varepsilon_F=(9\pi/8)^{2/3}\hbar^2/2mr_s^2$\,MeV.

\section{Construction of the folding matrix in a 1D spherical grid}
\label{seq:matrix1D}

The folding procedure for a 1D spherical representation involves the direct integration over the angles:
\begin{eqnarray*}
  \mathcal{F}_l(r,r')
  &\!=\!&
  \int d^2\Omega d^2\Omega' Y^*_{lm}(\Omega)
  \exp\left(-\frac{(\vec{r}-\vec{r}')^2}{\tilde{R}_{\mathcal{F}}^2}\right)
  Y^{\mbox{}}_{lm}(\Omega')
\\
  &\!=\!&
  2\pi\int d\eta
  \exp\left(-\frac{r^2+(r')^2-2rr'\eta}{\tilde{R}_{\mathcal{F}}^2}\right)
  P_l(\eta)
\\
&\!=\!&
2\exp\left(-\frac{r^2+ (r')^2}{\tilde{R}_{\mathcal{F}}^2}\right) i_{l}\left(\frac{2rr'}{\tilde{R}_{\mathcal{F}}^2}\right)
\end{eqnarray*}
where $\eta=\cos\vartheta$ and $i_{l}$ is the modified Bessel function of the first kind.

Alternatively, one can  carry out folding in the momentum representation by employing the Fourier-Bessel transformation:  
\begin{equation}
  \tilde{F}(\vec{p})
  \longrightarrow
  \tilde{F}_{lm}(p)
  =
  \int dr r^2\,d^2\Omega  j_l(pr)Y_{lm}(\Omega)
  \mathcal{F}(\vec{r}).
\end{equation}
The $p$ values are discretized to a set $p_\nu^{(l)}$ on a
finite $l$-dependent grid. 
The spatial folding can thus be represented as
\begin{equation}
  \mathcal{F}_l(r,r')
  =
  \sum_\nu j_l(p_\nu r)  
  \exp\left(-\frac{\tilde{R}_{\mathcal{F}}^2}{4}p_\nu^2\right)j_l(p_\nu r')\,.
\label{eq:Gauss1D}
\end{equation}

\section{Calibrated EDF parameters}
\label{sec:modelparams}

Table \ref{tab:model-params} shows the parameters for the Skyrme and the Fayans EDF with finite-range pairing at the minimum of the $\chi^2$ curves
in Fig.~\ref{fig:fit-width}.
The model parameters are complemented by the properties of symmetric nuclear matter at equilibrium (Table~\ref{Tab:NMproperties}). These help to characterize the basic physical properties of a functional.

\begin{table}[!htb]
\caption{Parameters of a Skyrme functional (left) and
  a Fayans functional (right) with finite-range pairing with $\tilde{R}_{\mathcal{F}}=0.7$\,fm and $\tilde{R}_{\mathcal{F}}=0.9$\,fm  respectively, treated in
  HFB and pairing cutoff energy $\varepsilon_\mathrm{cut}=15$\,MeV  at their
  minimum of the $\chi^2$ curve in Fig.~\ref{fig:fit-width}.  All Skyrme parameters are given in units of MeV for energies and fm
  for lengths. The model parameters for the Fayans functional are
  dimensionless by construction.}
\begin{center}
\begin{tabular}{lc}
%SV-min folding 0.7 newcut 15
\hline
\multicolumn{2}{c}{Skyrme}
\\
\hline
$C_0^{\rho}$ &  -756.4018463    
\\
$C_0^{\rho}$ &   430.0083542    
\\
$D_0^{\rho}$ &   847.1928481    
\\
$D_1^{\rho}$ &  -541.5625965    
\\
$C_0^{\Delta\rho}$&  -53.13268906    
\\
$C_1^{\Delta\rho}$&  -22.18923885    
\\
$C_0^{\tau}$&   6.218937572    
\\
$C_1^{\tau}$&   3.648292998    
\\
$C_0^{\nabla J}$&  -77.76109200    
\\
$C_1^{\nabla J}$&  -18.57579450    
\\
$\alpha$   &0.2745873647
\\
$V_{0,p}$&932.3973450
\\
$V_{0,n}$&899.6625795
\\
$\rho_\mathrm{pair}$&0.1987770272 
\\
\strut 
\\
\strut 
\\
\hline\\[-8pt]
\end{tabular}
$\quad\quad$
\begin{tabular}{lc}
\hline
\multicolumn{2}{c}{Fayans}
\\
\hline
  ${a_+^\mathrm{v}}$ & -9.608175465
\\
  ${h_{1+}^\mathrm{v}}$ &  0.6227052220
\\
  ${h_{2+}^\mathrm{v}}$ &  0.1726617465
\\
  ${a_-^\mathrm{v}}$ &  2.2427483150$\times10^{14}$
\\
  ${h_{1-}^\mathrm{v}}$ &  -2.316454053
\\
  ${h_{2-}^\mathrm{v}}$ &  5.2906966945$\times10^{14}$
\\
  ${a_+^\mathrm{s}}$ & 0.5483225953
\\
  $ h_{\nabla}^\mathrm{s}$ &  0.3511339663
\\
  $\kappa$ & 0.1906406585
\\
  $\kappa'$ &   -0.033380056233
\\
  $f_{\mathrm{ex},p}^\xi$ &   -8.270525940
\\
  $h_p^\xi$ &7.304858433
\\
  $h_{\nabla,p}^\xi$ &5.666948816
\\
  $f_{\mathrm{ex},n}^\xi$ & -11.28934445
\\
  $h_n^\xi$ &12.32075874
\\
  $h_{\nabla,n}^\xi$ &8.428702197
\\
\hline\\[-8pt]
\end{tabular}
\end{center}
%\end{minipage}

\label{tab:model-params}
\end{table}

\begin{table}[!htb]
\label{Tab:NMproperties}
\caption{The nuclear matter properties of the parametrization,
  equilibrium density $\rho_\mathrm{eq}$, binding energy per nucleon
  $E/A$, incompressibility $K$, isoscalar effective mass $m^*/m$,
  symmetry energy $J$, slope of symmetry energy $L$, and Thomas-Reiche-Kuhn sum rule
  enhancement $\kappa_\mathrm{TRK}$.}
% \begin{center}
\begin{tabular}{lrr}
%SV-min folding 0.7 newcut 15
\hline
 Property &Skyrme & Fayans 
 \\
 \hline
$\rho_\mathrm{eq}$ (fm$^{-3}$) &   0.1603 & 0.1633
\\
$E/A$ (MeV) &  -15.903 & -15.865
\\
$K$ (MeV) &   225.3 & 214.5
\\
$m^*/m$ &   0.9541 & 1.000
\\
$J$ (MeV) &   30.35 & 29.97
\\
$L$ (MeV) &    38.74 & 61.83
\\
$\kappa_\mathrm{TRK}$  &  0.0206 & 0.0000
\\
\hline
\end{tabular}
% \end{center}
%\end{minipage}

\end{table}

\bibliography{finrang-paper}
\end{document}